\documentclass{sig-alternate}

\usepackage{color}
\usepackage{amssymb,amsmath}
\usepackage{graphicx}
\usepackage{subfigure}
\usepackage{wrapfig}
\usepackage{epsfig}
\usepackage{endnotes}
\usepackage[normalem]{ulem}
\usepackage[hyphens]{url}
\usepackage{hyperref}
\usepackage{color}

\newenvironment{packed_itemize}{
\begin{itemize}
  \setlength{\itemsep}{1pt}
  \setlength{\parskip}{0pt}
  \setlength{\parsep}{0pt}
}{\end{itemize}}

\begin{document}

\conferenceinfo{TRUST}{'14 June 09 - June 11 2014, Edinburgh, United Kingdom}

\permission{}
\copyrtyr{}
\acmcopyr{}
\copyrightetc{}

\title{Experience report: community-driven reviewing and validation of publications}

\numberofauthors{2}

\author{
 \alignauthor
  Grigori Fursin\\ \vspace{0.2cm}
  \affaddr{INRIA and University of Paris-Sud, France}\\ \vspace{0.2cm}
  \email{grigori.fursin@inria.fr}
 \alignauthor
  Christophe Dubach\\ \vspace{0.2cm}
  \affaddr{University of Edinburgh, UK}\\ \vspace{0.2cm}
  \email{christophe.dubach@ed.ac.uk}
}

\maketitle

\begin{abstract}

In this report, we share our practical experience
on crowdsourcing evaluation of research artifacts and reviewing
of publications since 2008. We also briefly discuss encountered
problems including reproducibility of experimental results and
possible solutions.

\end{abstract}


\section{Crowdsourcing optimization of computer systems}

When trying to build a practical machine-learning based,
self-tuning compiler during the European FP7 MILEPOST project~\cite{milepost}
in 2006-2009, we faced multiple problems:
\vspace{-0.5em}
\begin{packed_itemize}

\item lack of common, large and diverse benchmarks and data sets
needed to build statistically meaningful predictive models;

\item dramatic lack of computational power to automatically
explore large program and architecture design and optimization
spaces required to effectively train compiler (building
predictive models);

\item difficulty to reproduce and validate already existing and
related techniques from existing publications due to a lack
of culture of sharing research artifacts and full experiment
specifications along with publications in computer engineering.

\end{packed_itemize}

Based on our background in physics and machine learning,
we proposed an alternative solution to develop a common
experimental infrastructure, repository and public web portal
that could help crowdsource program analysis and compiler
optimization across multiple users. Our goal was to persuade our
community to start sharing various benchmarks, data sets, tools,
predictive models together with experimental results along with 
their publications. This, in turn, could help the community validate and
improve past techniques or quickly prototype new ones using
shared code and data.

In the beginning, many academic researchers were not very
enthusiastic about this approach since it was breaking
``traditional'' research model in computer engineering where
promotion is often based on a number of publications rather than
on reproducibility and practicality of techniques or sharing
of research artifacts. Nevertheless, we decided to risk and
validate our approach with the community by releasing our whole
program and compiler optimization and learning infrastructure
together with all benchmarks and data sets. This infrastructure
was connected to a public repository of knowledge
(\url{cTuning.org}) allowing the community to share their
experimental results and consider program optimization as
a collaborative ``big data'' problem. At the same time, we shared
all experimental results as well as program, architecture and
data sets ``features'' or meta-information necessary for machine
learning and data mining together with generated predictive
models along with our open access publication
\cite{cm:29db2248aba45e59:a43178b20901ba15} ({\small
\url{http://hal.inria.fr/inria-00294704)}}.
As a result, we gained several important and practical experiences summarized below.

\paragraph{Use the community to test ideas}

The community served as a reviewer of our open access
publication, shared code and data, and experimental results 
on machine learning based self-tuning compiler. For example, 
our work was featured twice on the front page of {\small \url{slashdot.org}} 
news website with around 150 comments: \\
{\small 
\url{http://beta.slashdot.org/story/121289} \\
\url{http://beta.slashdot.org/story/103577}
} 
 
Of course, such public comments can be just "likes", "dislikes",
unrelated or possibly unfair which may be difficult to cope particularly 
since academic researchers often consider their work and publications unique 
and exceptional. On the other hand, quickly filtering comments and
focusing on constructive feedback or criticism helped us to
validate and improve our research techniques besides fixing obvious 
bugs. Furthermore, the community helped us find most relevant and
missing citations, related projects and tools - this is particularly
important nowadays with a growing number of publications, conferences,
workshops, journals, initiatives and only a few truly novel
ideas.

\paragraph{Engaging publicly is fun}

Exposing your research to a community and engaging
in public discussions can be really fun and motivating,
particularly after the following remark which we received
on Slashdot about MILEPOST GCC: "GCC goes online on the 2nd of
July, 2008. Human decisions are removed from compilation. GCC
begins to learn at a geometric rate. It becomes self-aware 2:14
AM, Eastern time, August 29th. In a panic, they try to pull the
plug. GCC Strikes back".

It is even more motivating to see that your shared techniques
have been immediately used in practice, improved by the
community, or even had an impact on industry. For example, our
community driven approach was referenced in 2009 by IBM for
speeding up development and optimization of embedded
systems~\cite{ctuning-press-release-ibm}, and in 2014 by Fujitsu
on "big data" driven optimizations for Exascale Computer
Systems~\cite{ctuning-press-release-fujitsu}.

\paragraph{Open access publications and artifacts bring us back to the root of academic research}

It is now possible to fight unfair or biased reviewing which
is sometimes intended to block other authors from publishing new
ideas and to keep monopoly on some research topics by several
large academic groups or companies. To some extent, rebuttals
were originally intended to solve this problem, but due to
an excessive amount of submissions and lack of reviewing time,
it nowadays has very little effect on the acceptance decision.
This problem often makes academic research looks like business
rather than collaborative science, puts off many students and
younger researchers, and was emphasized at all our organized
events and panels. 

However, with an open source publication and shared artifacts,
it is possible to have a time stamp on your open access
publication and immediately engage in public discussions thus
advertising and explaining your work or even collaboratively
improving it --- something what academic research was originally
about. At the same time, having an open access paper does not
prevent from publishing a considerably improved article in
a traditional journal while acknowledging all contributors
including engineers whose important work is often not even
recognized in academic research. For example, we received
an invitation to extend our open access paper on MILEPOST GCC and
publish it in a special issue of IJPP
journal~\cite{29db2248aba45e59:a31e374796869125}. Therefore, open
access and traditional publication models may possibly co-exist
while still helping academic researchers with a traditional
promotion.

\paragraph{Negative results should not be ignored}

It is even possible to share and discuss negative results
(failed techniques, model mispredictions, performance
degradations, unexplainable results) to prevent the community from making
the same mistakes and to collaboratively improve them. This
is largely ignored by our community and practically impossible
to publish currently.

In fact, negative results are in fact very important for machine learning based optimization and
auto-tuning. Such techniques are finally becoming popular
in computer engineering but require sharing of all benchmarks,
data sets and all model mispredictions --- besides positive results ---
to be able to improve them as it is already done in some
other scientific disciplines.

\paragraph{Sharing research artifacts brings people together and raises interest}

The community continue being interested in our projects mainly
because they are accompanied by all code and data enough
to reproduce, validate and extend our model-driven optimization
techniques. At the same time, sharing all research material in
a unified way helped us to bring interdisciplinary communities
together to explain performance anomalies, improve machine
learning models or find missing features for automatic program
and architecture optimization while treating it as a "big data"
problem. We also used it to conduct internal student competitions
to find the best performing predictive model. Finally, we used
such data to automatically generate interactive graphs
to simplify research in workgroups and to enable interactive
publications (as shown in the following online example:
\\ {\small \url{c-mind.org/interactive-graph-demo}}).

\paragraph{Reproducibility should not be forced but can come as a side effect}

Furthermore, such community driven research helped us to expose 
a major problem that makes reproducibility in computer engineering 
very challenging. We have to deal with ever changing hardware and software 
stack making it extremely difficult to describe experiments with all
software and hardware dependencies, and to explain unexpected
behavior of computer systems. Therefore, just reporting and sharing
experimental results including performance numbers, version of a compiler 
or operating systems and a platform is not enough - we need to preserve 
the whole experimental setup with all related artifacts and meta-information
describing all software and hardware dependencies.

\paragraph{Collective mind: a new beginning}

This problem motivated us to start developing a new methodology
and open source infrastructure (Collective Mind, {\small
\url{c-mind.org}}) to gradually describe, categorize, preserve
and share the whole experimental setups and all associated
research artifacts with their meta-description as public and
reusable components {\small \url{c-mind.org/repo}}, At the same
time, we and the community benefit from public discussions and
from agile development methodologies to continuously improve our
techniques and tools.

After many years of evangelizing collaborative and reproducible
research in computer engineering based on the presented practical
experience, we finally start seeing the change in mentality
in academia, industry and funding agencies. In our last ADAPT
workshop ({\small \url{adapt-workshop.org}} authors of two papers
(out of nine accepted) agreed to have their papers validated
by volunteers. Note that rather than enforcing specific validation rules,
we decided to ask authors to pack all their research artifacts
as they wish (for example, using a shared virtual machine or as
a standard archive) and describe their own validation procedure.
Thanks to our volunteers, experiments from these papers have been
validated, archives shared in our public repository ({\small
\url{c-mind.org/repo}}), and papers marked with a "validated
by the community" stamp.

\section{Towards a New publication and validation model}

Based on the above practical and mainly positive experiences, we 
propose a new publication model where preliminary evaluation 
of research artifacts and ideas is crowdsourced to an interdisciplinary
community using available web services.

\paragraph{Open access archives}

Papers should be submitted to open access archives such as arXiv
({\small \url{arxiv.org}}) and HAL ({\small
\url{hal.archives-ouvertes.fr}}).  All related artifacts (code,
data sets, models, experimental results) should be shared either
at authors' web pages or at some public sharing web services such
as ({\small \url{figshare.com}}). Finally, authors should submit
links to their open access papers and related artifacts to
a given workshop, conference or journal.

\paragraph{Public, threaded and ranked discussion forums}

After collecting all papers and related material, a new discussion
topic for each paper should be created at some social networking service 
with a ranking system such as slashdot, stackexchange or reddit.
Google+ or Facebook can be used too though just a number of "likes" 
or "dislikes" may be difficult to interpret. Authors will be strongly
encouraged to engage into public discussions about their work.

\paragraph{Program committee filter}

Small program committee should read discussions to quickly
get rid of publications where claims and experimental results
are obviously wrong, unreproducible or possibly plagiarized. Note, that
if authors disagree with the community (it may happen with too
novel or controversial ideas), their public arguments should help
pass this filter (similar to current rebuttal system). Remaining
papers should be sent to a specially selected and
possibly interdisciplinary committee based on topics of the submitted
papers as well as reviews (to address specific concerns from the
community).

Since we see a continuously growing number of papers
submitted to workshops, conferences and journals, such approach
can considerably reduce the burden of reviewers and help them focus
on possible issues already identified by the community while improving relevance
and quality of the reviews.

\paragraph{Final paper selection}

Program chairs can now select papers based on public discussions 
and professional reviews to ensure interesting and relevant
discussions at a workshop or conference. We believe that public 
discussions can also help avoid anonymous and unfair paper
rejections that are intended to keep monopoly on research. 
Finally, such approach can also help focus authors presentations on 
addressing questions and concerns raised during public discussions rather
than having long and formal introductions of the techniques.

\paragraph{Online open access journals}

Our approach transparently enables open access
journals --- we can now  immediately create online journal volumes
from the most interesting and highest ranked publications. At the
same time, existing and not necessarily open access journals can
also invite extended publications.

\paragraph{Co-existence with traditional publication models}

Note that we do not advocate to completely substitute "closed"
and professional reviewing at current conferences and journals.
Neither do we advocate for an open access to all
publications and research artifacts - it is normal if a company
or a researcher would like to possibly patent and
commercialize their work while still presenting it to the public.
In such case, we still need to allow traditional submission
to workshops, conferences or journals along with public submissions.
However, we would still like to validate experimental results
in these publications. In such cases, we may even envision that some 
committee members may need to sign NDA to validate such experimental results.

Therefore, our publication model can easily co-exist with current
models instead of trying to abruptly substitute them.
Furthermore, we hope that it will reduce reviewers' burden,
improve quality and fairness of the reviews and will restore
attractiveness of academic research in computer engineering as
a traditional, collaborative and fair science rather than
hacking, publication machine or monopolized business.

\section{Future work}

We plan to validate the presented publication model at our next ADAPT
workshop (\url{adapt-workshop.org}) and possibly at some existing
conferences and journals in computer engineering.

Note, that experimental reproducibility comes naturally in our
publication model as a side effect rather that only because it is
a noble thing to do. However, we continue experiencing
considerable difficulties when reproducing complex experimental
setups from existing publications often due to specific
requirements placed on operating systems, libraries, benchmarks,
data sets, compilers, architecture simulators, and other tools,
or due to a lack of precise specifications, lack of all software
dependencies, and lack of access to some hardware.

Similar problems with reproducibility were also recently 
reported in several other related initiatives on validating experimental
results~\cite{project-aec,project-repro-cs,project-occam}.
Therefore, we decided to join together during the ACM SIGPLAN TRUST
workshop~\cite{trust14} to discuss
technological aspects of enabling collaborative and reproducible
research and experimentation particularly on program and
architecture empirical analysis, optimization, simulation,
co-design and run-time adaptation including how to:

\begin{packed_itemize}
\item capture, catalog, systematize, modify, replay and exchange
experiments (possibly whole setups with all artifacts including
benchmarks, codelets, datasets, tools, models, etc);
\item validate and verify experimental results;
\item deal with a rising amount of experimental data using
statistical analysis, data mining, predictive modeling, etc.
\end{packed_itemize}

Finally, we will continue investigating frameworks and
repositories to share the whole experimental setups with related
artifacts and their meta-description including all software and
hardware dependencies such as CARE ({\small
\url{reproducible.io}}), CDE ({\small
\url{www.pgbovine.net/cde.html}}), Docker ({\small
\url{www.docker.io}}), IPython Notebook ({\small
\url{ipython.org/notebook.html}}), Collective Mind
({\small \url{c-mind.org}}), and many others.

\section{Acknowledgments}

We would like to thank all the participants of our events (BOFs,
workshops, thematic sessions and panels) on collaborative and
reproducible research and experimentation in computer engineering
since 2008 including Jack Davidson, Lieven Eeckhout, Sascha
Hunold, Anton Lokhmotov, Alex K. Jones, Bruce Childers, Daniel
Mosse, Vittorio Zaccaria, Christian Bertin, Christophe Guillon,
Christoph Reichenbach, Marisa Gil, Lasse Natvig, David Whalley,
Cristina Silvano, Steve Furber, Paul Kelly, Thomas Wenisch,
Davide del Vento, Jean Luc Gaudiot as well as cTuning and HiPEAC 
communities for interesting discussions and feedback.

\bibliographystyle{plain}
\bibliography{article_hal}

\end{document}